\RequirePackage{lineno}

\documentclass[aps,prb,twocolumn,groupedaddress,superscriptaddress]{revtex4}

\usepackage{graphicx}
\usepackage{amssymb}

\begin{document}



\title{Non-simple flow behavior in a polar van der Waals liquid: structural relaxation under scope}


\author{S. Arrese-Igor}
\affiliation{Centro de de F\'{\i}sica de Materiales (MPC), Centro Mixto CSIC-UPV/EHU, 
Paseo Manuel Lardizabal 5, 20018 San Sebasti\'an, Spain}

\author{A. Alegr\'{\i}a}
\affiliation{Centro de de F\'{\i}sica de Materiales (MPC), Centro Mixto CSIC-UPV/EHU, 
Paseo Manuel Lardizabal 5, 20018 San Sebasti\'an, Spain}
\affiliation{Departamento de Pol\'{\i}meros y Materiales Avanzados UPV/EHU, 
Apartado 1072, 20080 San Sebasti\'an, Spain}

\author{J. Colmenero}
\affiliation{Centro de de F\'{\i}sica de Materiales (MPC), Centro Mixto CSIC-UPV/EHU, 
Paseo Manuel Lardizabal 5, 20018 San Sebasti\'an, Spain}
\affiliation{Departamento de Pol\'{\i}meros y Materiales Avanzados UPV/EHU, 
Apartado 1072, 20080 San Sebasti\'an, Spain}
\affiliation {Donostia International Physics Center, 
Paseo Manuel Lardizabal 4, 20018 San Sebasti\'an, Spain}



\date{\today}

\begin{abstract}
The non-exponential character of the structural relaxation is considered one of the hallmarks of the glassy dynamics and in this context, the singular shape observed by dielectric techniques has attracted the attention of the community for long time. Particularly, the exceptionally narrow dielectric response of polar glass formers has been attributed so far to dipolar cross correlations. Here we show that dipole interactions can couple to shear stress and modify the flow behavior preventing the occurrence of the simple liquid behavior. We discuss our findings in the general framework of the glassy dynamics and the role of intermolecular interactions. 
\end{abstract}


\maketitle



Liquids possess the ability to avoid crystallization upon cooling -if the cooling rate is just high enough- and become a solid glass in a phenomenon known as the glass transition. When approaching the glass transition temperature the dynamics of the supercooled liquid becomes exponentially slower till eventually the system solidifies as a consequence of the arrest of the molecular motions leading to the so called structural $\alpha$ relaxation \cite{Tg,Tg2}. The broad diversity in structural, chemical and physical properties of the materials capable of forming a glass complicate a full understanding of the glass transition problem. Notwithstanding, substantial progress was made in elucidating common generic features for the structural $\alpha$ relaxation \cite{Fruitfly}. In this quest for universal characteristics however, it is important to properly identify those molecular motions whose arrest ultimately lead to the glass transition, since depending on the particular class of materials these may coexist with other material specific dynamics.
As a consequence, in practice, unambiguous determination of the characteristics of the $\alpha$ relaxation is not straightforward due to possible overlapped contributions from other processes which may apparently rule-out some $actual$ universal behavior \cite{RosslerBetasWings,GainaruNMRshape}.

The nonexponential shape of the structural $\alpha$ relaxation is one of the hallmarks of glassy dynamics, many works pointing to a common universal shape for this process \cite{KNissJCP2009, BlochoDLSshape, GainaruNMRshape, GainaruShearshape0, GainaruShearshape}. A singular behavior is found however in this respect for polar systems as seen by dielectric techniques \cite{PaluchPolarBeta}. The narrower lineshape observed in this case has been ascribed to dipolar interaction effects \cite{PaluchPolarBeta,BlochoTBP, PaluchMDS2022}, though details on the microscopic origin and influence on the structural relaxation though still remain unclear.
Under the assumption that depolarized dynamic light scattering (DDLS) measurements render the $\alpha$ relaxation, Pabst and co-workers exploited the differences observed between the dielectric and DDLS data to identify an additional slow contribution in the dielectric spectra of a series of phenyl monoalcohols, glycerol polyalcohol and tributyl phospate (TBP) polar van der Waals liquid \cite{BlochoGC,BlochoPh,BlochoTBP}. The presence of a slow Debye-like relaxation in addition to the $\alpha$ one is a well-known feature in many monoalcohols and it is commonly ascribed to the relaxation of hydrogen bond mediated supramolecular structures \cite{100years}. However, although often addressed as reminiscent of the Debye
process in monoalcohols, for other substances\cite{CorreiaIBUP,PaluchPTZ,PaluchIBUP,GainaruTBP,BlochoGC,BlochoTBP} we still lack a full understanding of the molecular origin of this slow Debye-like contribution.  Pabst et al. rationalized differences in DDLS and dielectric data proposing the presence of an underlying and universal shape {\it self} part (which they denoted $\alpha$ relaxation) and a dominant slow $cross$ contribution in the dielectric response of glycerol and TBP. Supporting this view, recent molecular dynamics simulations on model liquids show that $cross$-terms are slower than and dominate over {\it self}-terms on the decay of the total dipole moment correlation function for strongly polar liquids \cite{PaluchMDS2022}.

According to the general believe, vitrification, relaxation of the structure in the glassy state (structural recovery) and the equillibrium dynamics above but close to the glass transition are collective phenomena. As a consequence, the idea that only the faster {\it self}  or autocorrelation contribution can render the structural dynamic evolution was recently questioned  \cite{GainaruTBP}. Looking at the ageing behavior of TBP alkyl phosphate, Moch et al. stated that the same collective dynamics governs the molecular flow, the structural recovery and the dielectric response, claiming a crucial role in the structural dynamics for the $cross$ correlation effects dominating the dielectric spectra (Debye-like term), against the single particle contribution identified by DDLS. Calorimetric studies of monoalcohols on the other hand, indicate that their Debye relaxation is not involved in the thermal glass transition \cite{RichertAC-dsc_2E1H, RichertACdsc_2OH, RichertJPCL2020}.

All the mentioned works put under the scope the question of the nature and origin of the 'structural relaxation'.  Over several decades, many efforts have been devoted to the understanding of the vitrification phenomenon
and, for this, to the characterization of the dynamics of the associated $\alpha$ relaxation in glass-forming systems of diverse nature and by different experimental techniques. 
Probably influenced by the Mode Coupling Theory \cite{MCT,ReviewMCT}, the term 'structural relaxation' has been used since the 1990$^{'}$s to refer to the mechanism leading to the decay of the coherent scattering function at the main structure factor peak, revealing thereby the time dependence of inter-molecular correlations. The presence of different kind and degree of specific non-covalent interactions though can potentially modify the dynamic response of liquids. Moreover, ionic or hydrogen bonding for example, are known to produce mesoscopic structures in some cases \cite{Tomsic2007,Lehtola2010,Perera2017}, so that additional dynamics could emerge related to the formation and relaxation of the mentioned mesoscopic structures. 
Under such situations, the concept behind 'structural relaxation' is much broader and could be applied to diverse mechanisms leading to the decay of density fluctuations at different length scales, including those involving specific interactions and  mesoscopic structures. The question is now, what dynamics do we regard as $\alpha$ relaxation? those contributing to the vitrification? to structural recovery? to fluctuations of the enthalpy at equilibrium? to which extent do collective dynamics at intermediate timescales contribute to the processes mentioned above? 
These are not trivial questions at all. In the absence of additional complexity introduced by non-covalent interactions or other peculiarities, terms like $\alpha$ relaxation, structural relaxation, glass-transition dynamics and structural recovery have often been indistinctly used. At the light of the last findings, \cite{RichertRelaxRecov}, however , it seems that the scientific community will need to clarify the universal or specific (interaction dependent) nature of the new emerging processes and their role on the vitrification and the structural recovery phenomena. 

In this regard, we present here for the first time evidence of the bimodal shear response of a polar van der Waals liquid, TBP.  The data presented herein demonstrate that dipolar interactions are capable of modifying the rheologic response of a liquid preventing the occurrence of the simple liquid behavior often assumed for many low molecular weight glass forming systems. We discuss and compare TBP results with the dynamic behavior observed for hydrogen bonded systems (mono and polyalcohols) and other polar and non polar glass formers, allowing us to introduce new perspective on the phenomenology and role of specific non-covalent interactions in the structural relaxation of glass forming systems.

\begin{figure}[!htb]
\begin{center}
\includegraphics[width=.7\columnwidth]{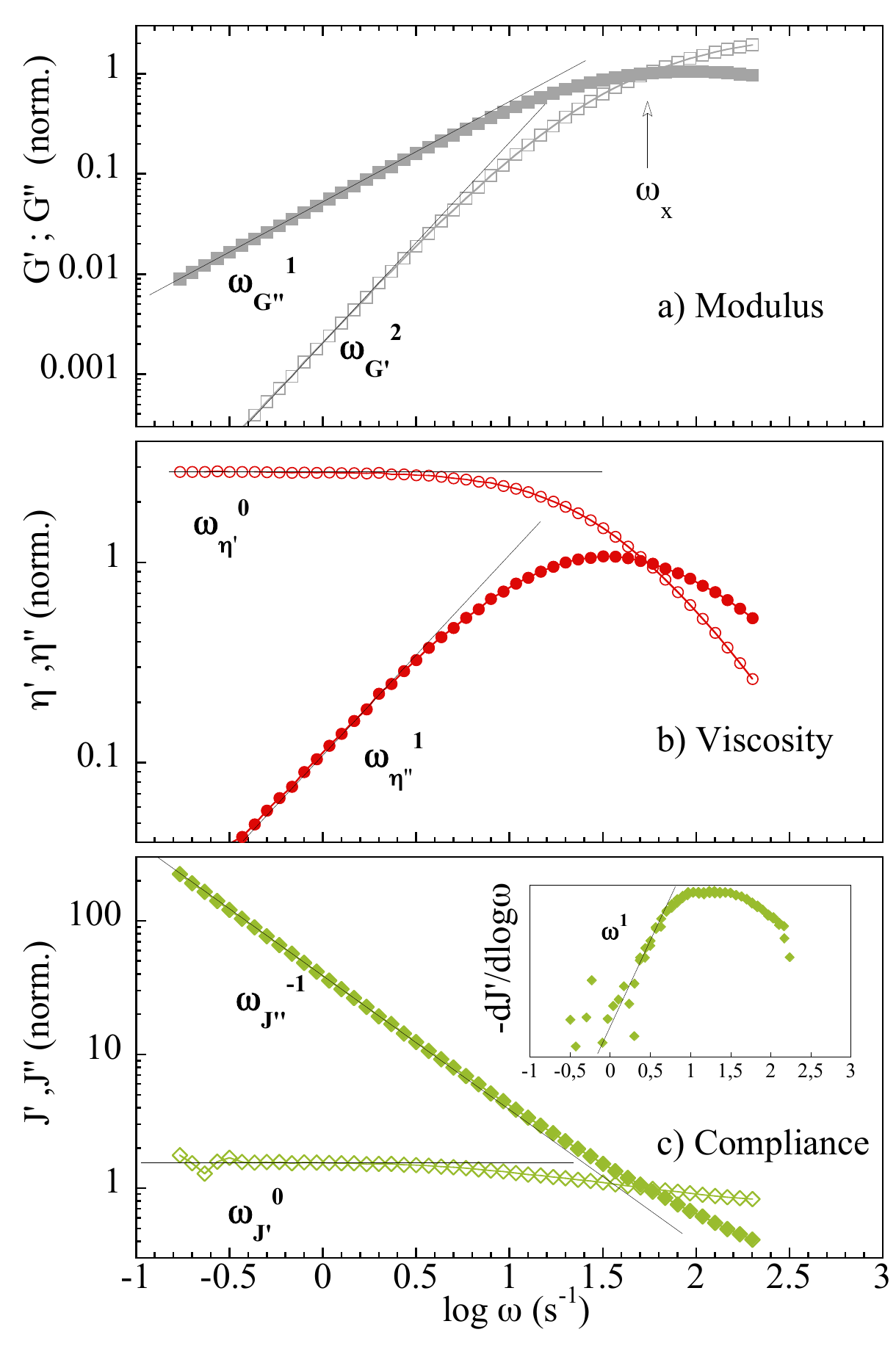}
\end{center}
\caption{Shear relaxation of TBP at 146K: (a) modulus, (b)viscosity and (c) compliance. Data in the y-axis was normalized to the crossing value of the real and complex components. Lines are guides for the eye representing different power law behaviors. The inset in (c) represents the derivative of $J'$.}
\label{MEC}
\end{figure}

The dielectric and shear response of glass forming liquids in general is characterized by the presence of the so called  $\alpha$ relaxation \footnote{Depending on the nature of the glass former, additional processes may also be present as secondary relaxations, the normal mode characteristic of chain relaxation in polymers, or the Debye-like relaxation related to supramolecular hydrogen bond aggregates in monoalcohols, for example}.
The simplest approach to describe the viscoelasticity of glass forming liquids is the Maxwell model, $G^{*}(\omega)=G_{p}/(1-i/\omega\tau_{M})$.
At high frequencies, the crossing point of the real and imaginary modulus $G'(\omega_{x})=G''(\omega_{x})$ marks a timescale for the $\alpha$ relaxation, while at lower frequencies $G'(\omega)$ and $G''(\omega)$ show the so called terminal behavior, ($G'(\omega)\propto \omega^{2}$ and $G''(\omega)\propto \omega$) where the liquid presents pure viscous flow.
Simple liquids closely satisfy this model, and in practice, the non-exponential nature of the $\alpha$ relaxation has little impact in the low frequency flank of the shear modulus, so that pure viscous flow is shortly established after the crossing frequency $\omega_{x}$ \cite{Vademecum}. The separation between $\omega_{x}$ and the crossover frequency to terminal behavior $\omega_{t}$ can be used then as a measure or indication of the $'simplicity'$ of a certain rheologic response. Systematic comparison of the separation between these two characteristic features for different glass forming liquids allowed to reveal subtle effects of hydrogen bonding on the shear response of polyalcohols for example \cite{MioGC}. 

\begin{figure}[!htb]
\begin{center}
\includegraphics[width=0.7\columnwidth]{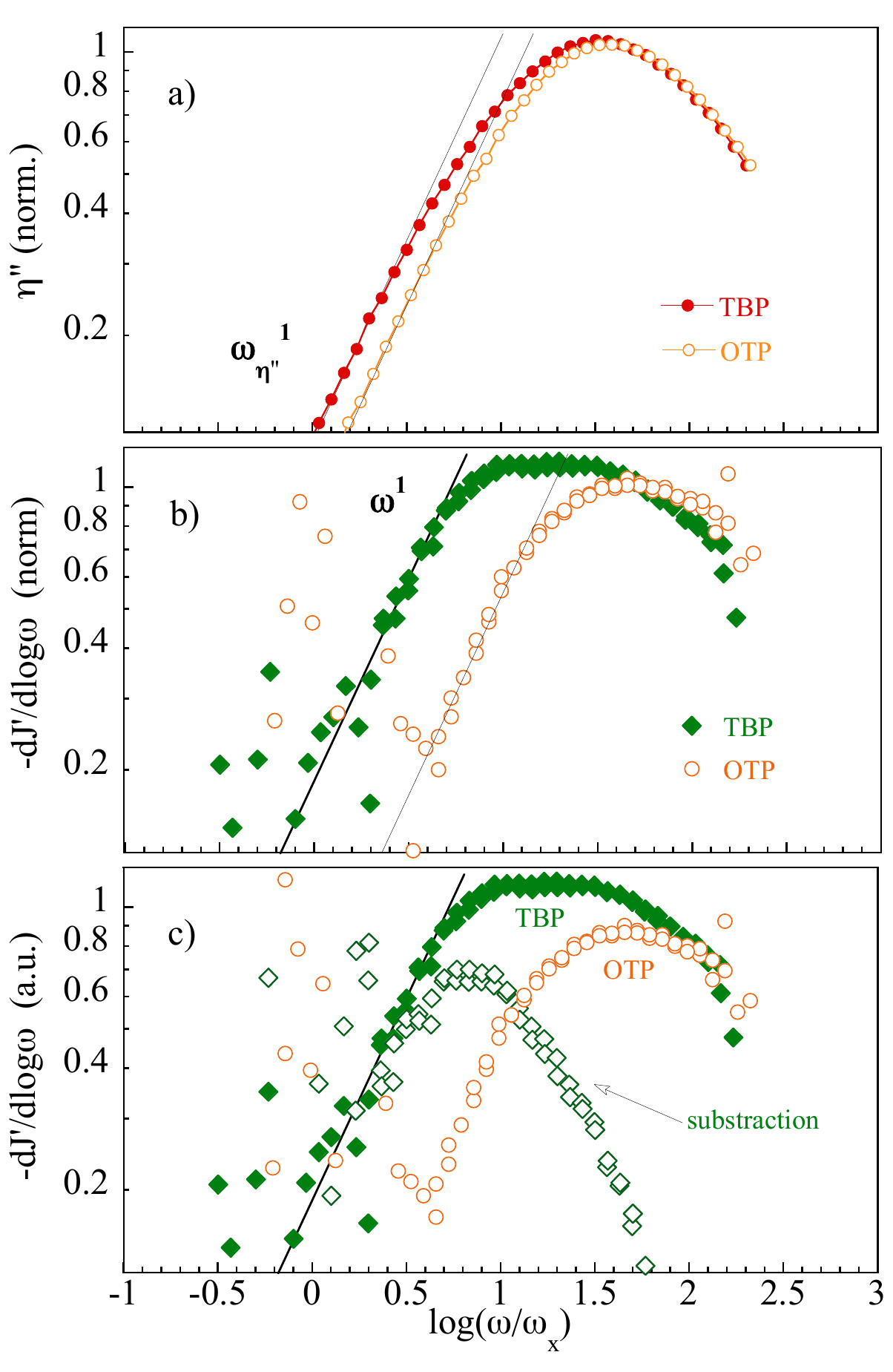}
\end{center}
\caption{Comparison of the shear relaxation of TBP and OTP. Lines are guides for the eye representing $\omega^{1}$ power law.}
\label{Compli}
\end{figure}

Figure \ref{MEC} shows the shear relaxation of TBP at 146K. Although at first sight the shear response of TBP seems to be relatively simple, close inspection reveals that the onset of the terminal behavior sets in more than a decade below the crossing frequency $\omega_{x}$. This behavior is clearly different from that observed for other simple liquids \cite{Vademecum, MioGC} either indicating non-universal shape for the structural relaxation of TBP relative to other simple liquids, or the presence of some additional slow contribution. 
Figure \ref{Compli} compares the shear response of TBP with that of $o$-tertphenyl (OTP), which is representative of the behavior of simple liquids \cite{RichertOTP}. In order to compare the shape of the relaxation, the frequency axis was normalized to $\omega_{x}$ for the shear modulus, and the y-axis to the value of the depicted magnitude at $\omega_{x}$. As it can be seen in panel a), both samples show almost indistinguishable shape for the viscosity losses at high frequencies but when decreasing frequency, OTP reaches terminal behavior at higher frequencies than TBP. In  panels b) and c) data are represented as the derivative of the real shear compliance. Interestingly, it turns out that slower contributions are relatively enlarged in this representation and as a consequence, the difference between OTP simple liquid and TBP becomes even more evident. This together with the relatively flat and broad peak shape of the TBP response in this representation strongly suggest the presence of an underlying additional low frequency contribution. In order to further check the consistency of this idea, panel c) shows a tentative determination of a slower contribution (empty diamonds), which was obtained after subtracting from TBP data those of OTP rescaled in the y-axis to match the high frequency tail of TBP data.
With all, detailed analysis of the shear response presented herein evidences that TBP does not exhibit a simple rheological response. To our knowledge, this is the first time that such behaviour is reported for the molecular flow of a 'simple' van del Waals liquid.  The comparison of these results with dielectric relaxation behavior (see below) indicates that dipolar interactions are capable as well of modifying the shear response of a liquid preventing the occurrence of the simple liquid behavior often assumed for many low molecular weight glass forming systems.

\begin{figure}[!htb]
\begin{center}
\includegraphics[width=0.9 \columnwidth]{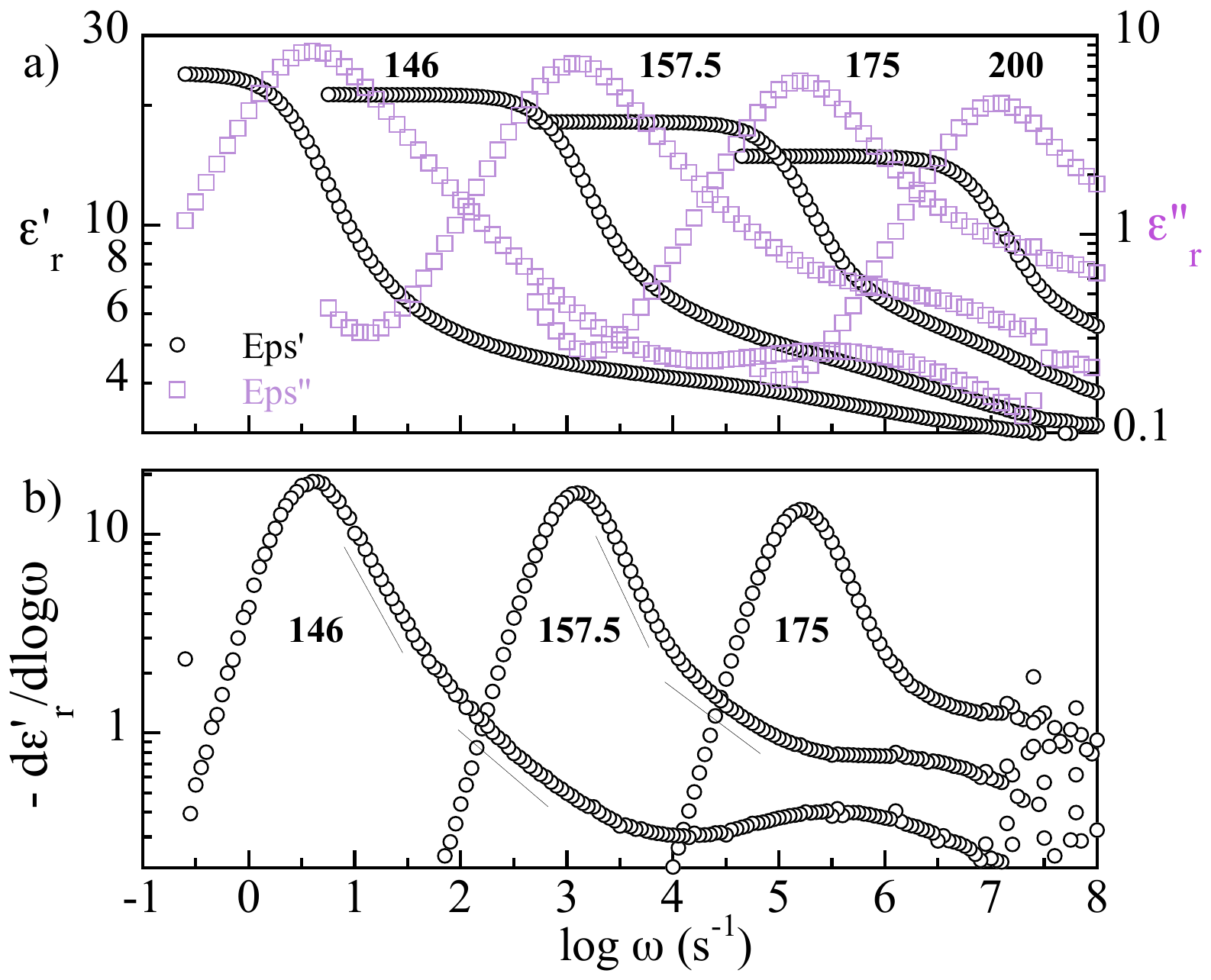}
\end{center}
\caption{Dielectric response of TBP: (a) $\epsilon'_{r}$ ($\circ$ left axis) and $\epsilon''_{r}$ ($\square$ right axis). (b) derivative of $\epsilon'_{r}$ ($\circ$).}
\label{BDS}
\end{figure}

Figure \ref{BDS} shows the dielectric response of TBP for several temperatures (see supporting information for details). At first sight two different relaxations could be identified: a prominent low frequency peak and a faster and less intense secondary relaxation. When representing data as the derivative of $\epsilon '$ the high frequency wing of the main relaxation clearly shows a change of power law behavior (see small solid lines in figure \ref{BDS}b ) which could be suggestive of the presence of an underlying contribution. Note that although such effects are quite evident in the derivative representation they are far more subtle in $\epsilon''(\omega)$ data representation and can easily go unnoticed for the naked eye. 
The comparison of BDS and shear data at 146K shows that the frequency at $\eta^{''}$ maximum and $\omega_{x}$ locate at the high frequency wing of the slower dielectric peak, nearby the change of slope referenced in figure \ref{BDS}. The maxima of the dielectric relaxation on the other hand, is more than one decade slower than $\omega_{x}$. This result is at odds with systematic studies showing that the frequency of the dielectric $\alpha$ relaxation represented as permittivity locates close to $\omega_{x}$ and is strongly correlated with it \cite{Mecanicas, Vademecum, MioGC}.  When the frequency axis of the dielectric permittivity data of different materials (polymers, monoalcohols and low molecular weight liquids) is normalized with respect to their own shear $\omega_{x}$, the position of the maxima of all the resolved dielectric $\alpha$ peaks perfectly match, evidencing the mentioned strong correlation between this two events \cite{MioGC}. 
The fact that the dielectric relaxation of TBP does not conform to the mentioned correlation together with the hints of an underlaying contribution in the high frequency wing of the main dielectric peak, strongly support the idea of overlapped contributions in the dielectric signal too. Consistently then, and following the behavior observed for other glass forming systems, a fast dielectric component would locate at the high frequency wing of the main dielectric peak of TBP, more or less coinciding with the maximum observed in $\eta''(\omega)$. The frequency of the dielectric maxima dominated by the slower contribution, on the other hand, matches quite well the crossover frequency to terminal behavior, $\omega_{t}$. While this match may seem a coincidence, it is systematically fulfilled as well for the dielectric Debye relaxation of monoalcohols and their respective crossover frequency to pure viscous flow \cite{Mecanicas}. 

\begin{figure}[!htb]
\begin{center}
\includegraphics[width=.9\columnwidth]{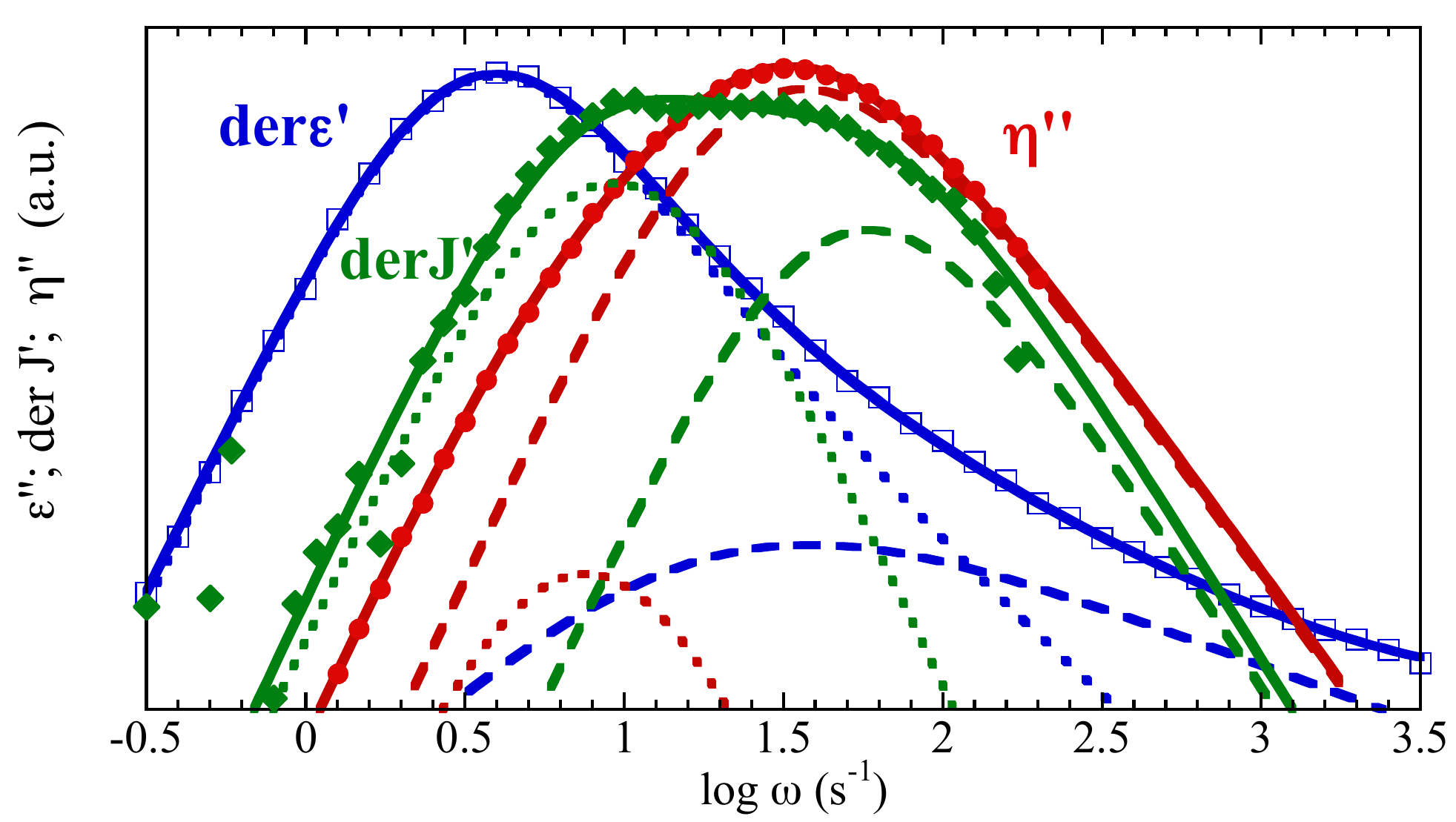}
\end{center}
\caption{Decomposition of shear and dielectric data into two contributions. Symbols represent experimental points:  $\eta''$ - red $\circ$ -; derivative of t $J'$ - green $\diamond$ -;  derivative of the $\epsilon'_{r}$ - blue $\square$ -.  For each magnitude (color) solid line stands for model curve whereas dotted and dashed lines represent the slow and fast components of the fitting, respectively.} 
\label{fittings}
\end{figure}

All together, the presence of two different contributions to the relaxation dynamics of TBP is consistent and evident for shear and BDS data.  Figure \ref{fittings} shows the phenomenological decomposition of the shear and BDS data in two components (see suplementary for details). For shear data, the weight of each contribution varies depending on the represented magnitude but the timescales of the various components reasonably agree for compliance and viscosity representations. Regarding the dielectric signal, the fast dielectric component agrees fairly well with those obtained from shear data. The low frequency dielectric component however, is around 0.5 decades slower than the slower shear component. Interestingly, the comparison of the shear and dielectric response of several monoalcohols shows analogous phenomenology, where the maxima of the dielectric Debye relaxation locates around 0.5 decades below the slowest component of the shear relaxation \cite{MioGC,Vademecum}. 
Microscopic molecular motions can couple in a different way to the various relaxing experimental observables so that in principle, timescales for different correlators do not need to coincide \cite{BirteTimescaleCoupling}. Nevertheless, as commented before, the timescales of the dielectric permittivity and those of shear relaxation are highly correlated for the $\alpha$ relaxation. For the slower component, a similar correlation between the timescales observed by both techniques seems to prevail, but involving a higher shift of about 0.5 decades. The distinct magnitude of the separation between shear and dielectric times for different processes but not for different systems could be then related to the microscopic nature of the motions involved, as when comparing the timescales for different Legendre polynomial correlation functions. The systematic study of the shift between the timescales observed by various techniques may thus in the future aid in resolving the microscopic origin of the relaxations.

The presence of two different contributions in the dielectric spectra of TBP was also previously proposed by Pabst et al. \cite{BlochoTBP} after subtracting from the overall dielectric signal the {\it self}  relaxation component determined by DDLS. Even though our results and those in reference \cite{BlochoTBP} are qualitatively consistent, there is not a clear quantitative agreement between the various timescales (see supplementary).
In this regard we want to stress the fact that, shear relaxation experiments by their own and devoid of any assumption, scaling or comparison, are sensible to the presence of two contributions. Regarding dielectric measurements, although not in an unquestionable way, we have seen that the derivative representation of the permittivity data by its own already suggest the presence of an underlying process which went unnoticed in the literature for a while.

Proper understanding and analysis of the data as obtained herein is crucial to comprehend and unify various results in the literature covering questions related to the shape of the $\alpha$ relaxation, its universality, the cooperative nature of the microscopic motions behind and how these are reflected in various experimental techniques. As we have shown above, within our analysis dielectric and shear relaxation show pretty much the same picture. The work of Moch et al. \cite{GainaruTBP} did not contemplate the multimodal character of the shear response of TBP. As a consequence, timescales extracted from shear compliance and viscosimetry data were fitted together under a single Vogel Fulcher Tamman equation \cite{VFT1,VFT2} and showed different temperature dependence from dielectric times (figure S2 in reference \cite{GainaruTBP}).
As shown above, shear compliance measurements underline the slower component while viscosity representation emphasizes the faster one. As a consequence, in agreement with our results and data analysis shear compliance times by Moch et al. \cite{GainaruTBP} closely follow slow component timescales while viscosimetry data reflect the faster one. Wu and coworkers studied vitrification dynamics of TBP at various cooling rates by differential scanning calorimetry and found that the fragility index and the stretching exponent for the extracted timescales differed from those found by dielectric spectroscopy \cite{NgaiTBPfragility} after analyzing the main relaxation as a single process. If we take into account however the existence of two components, the stretching parameter and the fragility index for the faster dielectric component obtained in our analysis are pretty much compatible with those derived from calorimetric measurements.  
According to these results one might follow that the faster component could correspond to the $\alpha$ relaxation as the principal dynamics contributing to the enthalpic fluctuations in equilibrium. However, the role of the slower component in the relaxation of the structure still can not be completely ruled out. In the above mentioned work of Moch et al.,\cite{GainaruTBP} physical aging experiments on TBP at 137K showed that structural recovery dynamics can be well predicted on the basis of equilibrium timescales measured by dielectric techniques (and dominated by the slow component) \cite{GainaruTBP}. Model analysis of dielectric data in terms of two components suggest that the timescale separation between the two components would decrease when approaching the glass transition, so it could also be the case that both dynamics merge and become indistinguishable at such low temperature as 137K.

It is remarkable that the dielectric and shear response of TBP and glycerol are highly analogous \cite{MioGC,Vademecum}. Moreover, in both cases calorimetric response seems to reflect cross-correlation dynamics or the slow component as seen by different techniques (and dominating the dielectric signal)\cite{RichertJPCL2020,GainaruTBP}. In this scenario, the hydrogen bond switch and the dynamic interchange of dipolar-dipolar interactions between different molecules would then play the same role on the relaxation of the structure. In the case of monoalcohols though, the slowest Debye contribution does not reflect this individual hydrogen bond switch, but the relaxation of the whole supramolecular chain structure \cite{GainaruNMRBuOH}. The hydrogen bond exchange in monoalcohols has been reported to have an intermediate timescale between those of the Debye and the $\alpha$ relaxation \cite{GainaruNMRBuOH,GainaruNMR2E1H,Mecanicas,Sales}. Interestingly, calorimetric studies on monoalcohols indicate that their Debye relaxation is not involved in the thermal glass transition \cite{RichertAC-dsc_2E1H, RichertACdsc_2OH, RichertJPCL2020}. On the other hand, it is worth notting that not all polar glass formers with narrow dielectric response seem to exhibit 'dynamically decoupled' (within experimental uncertainty) $cross$ and {\it self} correlations so as to produce a double step decay of the shear response. As an example, propylene carbonate is a quite polar glass forming system showing simple liquid behavior analogous to that of OTP, and for which dielectric timescales match shear $\omega_{x}$\cite{MioGC}. Therefore, we should not directly assume that all the low frequency Debye-like contributions have the same molecular origin even if all of them are experimentally well-captured by dynamic correlators including intermolecular cross correlations.

In the future, it will be necessary to deal with the relaxation of the structure at different levels, and elucidate which of those processes take part or not on the vitrification mechanisms, the structural recovery or the equilibrium dynamics of supercooled liquids near the glass transition temperature. In this sense, differential scanning calorimetry and the various techniques and protocols developed to access different aspects of the glass transition may be of great interest.

In summary, detailed analysis of the shear response presented herein evidences that TBP does not exhibit a simple rheological response. To our knowledge, this is the first time that such behavior is reported for the molecular flow of a 'simple' van del Waals liquid. The shear response of TBP is more extended in frequency than that of a Maxwell relaxation and the crossover to the terminal behavior sets in at lower frequencies than expected for a simple liquid. Moreover, the weight of the different contributions vary among the various representations of the shear response underlaying in a complementary way the various dynamic processes and making more evident the multimodal response of the liquid. With all, the data and analysis presented herein indicate that dipolar interactions are capable as well of modifying the shear response of a liquid preventing the occurrence of the simple liquid behavior often assumed for many low molecular weight glass forming systems.


\section*{Sample CRediT author statement}\textbf{S. Arrese-Igor} Conceptualization, Methodology, Formal analysis, Investigation, Writing - Original Draft, Visualization. \textbf{A. Alegr\'{\i}a} Data curation, Resources, Writing - Review $\&$ Editing, Supervision, Project administration, Funding acquisition.  \textbf{J. Colmenero} Resources, Writing - Review $\&$ Editing, Supervision, Project administration, Funding acquisition.

\begin{acknowledgments}
We acknowledge the Grant PID2021-123438NB-I00 funded by MCIN/AEI /10.13039/501100011033 and by 'ERDF A way of making Europe' and grant IT1566-22 by the Basque Government.
\end{acknowledgments}


\end{document}